\documentclass[twocolumn,aps]{revtex4}
\usepackage{epsfig,amsmath}

\def\d{{\rm d}}

\def\e{{\rm e}}
\def\i{{\rm i}}

\def\zhatc{\hat Z_c}
\def\zhatb{\hat Z_b}
\def\zhatf{\hat Z_f}
\def\cbar{\bar {\rm c}}
\def\bbar{\bar {\rm b}}
\def\ccb{{\rm c}\bar{\rm c}}
\def\bbb{{\rm b}\bar{\rm b}}
\def\nuc{\nu_c}
\def\nucb{\nu_{\bar c}}
\def\nub{\nu_b}
\def\nubb{\nu_{\bar b}}
\def\nuf{\nu_f}
\def\nufb{\nu_{\bar f}}
\def\nucj{\nu_{cj}}
\def\nucbj{\nu_{\bar c j}}
\def\nubj{\nu_{bj}}
\def\nubbj{\nu_{\bar b j}}
\def\nufj{\nu_{fj}}
\def\nufbj{\nu_{\bar f j}}
\def\scc{\sigma_{c \bar c}}
\def\sbb{\sigma_{b \bar b}}
\def\sff{\sigma_{f \bar f}}
\def\ppb{{\rm p}\bar{\rm p}}

\begin{document}

\title{Production of multiply heavy flavoured baryons from Quark
Gluon Plasma in relativistic heavy ion collisions} 

\author{F. Becattini}\affiliation{Universit\`a di Firenze}
\affiliation{INFN Sezione di Firenze, Florence, Italy} 

\begin{abstract}
It is argued that in heavy ion collisions at LHC there could be a 
sizeable production of baryons containing two or three heavy 
quarks from statistical coalescence. 
This production mechanism is peculiar of Quark Gluon Plasma and the
predicted rates, in heavy ion collisions at LHC energy, exceed 
those from a purely hadronic scenario, particularly for $\Xi_{bc}$ 
and $\Omega_{ccc}$. Thus, besides the interest in the discovery of 
these new states, enhanced ratios of these baryons over singly heavy
flavoured hadrons, like B or D, in heavy ion collisions with respect
to pp at the same energy, would be a clear indication of kinetic 
equilibration of heavy quarks in the Quark Gluon Plasma.
\end{abstract}

\maketitle

Charm and bottom quarks are expected to be abundantly produced in hadronic
collisions at very high energies. In heavy ion collisions (HIC), multiple pair 
production is expected to occur, with average multiplicities which, at the LHC 
energy of 5.5 TeV, attain ${\cal O}(10)$ for bottom and ${\cal O}(100)$ for charm 
in central collisions. These quarks, produced in the early 
stage of the collision off hard scatterings, lose energy in the Quark 
Gluon Plasma (QGP) and, if the lifetime of the source is long enough, may reach 
thermal \cite{rapp} (not chemical, as their annihilation rate is very low) 
equilibrium within the medium. 
At the hadronization point, they will coalesce into hadrons. If coalescence process
occurs statistically at the hadronization temperature, there is a finite chance 
that two or even three of them coalesce into the same particle, thus giving rise 
to multiply heavy flavoured hadrons, particularly baryons. This phenomenon is 
likely to occur only if heavy quarks get very close to thermal equilibrium 
reshuffling over a large region because high momentum quarks, most likely, will 
hadronize into different particles unless two or three of them emerge very close 
in momentum from the hard process. The latter production 
mechanism, where multiply heavy flavoured hadrons arise from {\em correlated} quarks, 
predicts multiplicities which, at some large energy, are exceeded by those predicted
by coalescence of {\em uncorrelated} quarks. The ultimate reason of this effect is 
that the average multiplicity of heavy quark pairs increases
faster than soft hadrons multiplicity as a function of centre-of-mass energy.
This is in turn related to the volume of the system at freeze-out. Consequently, 
the system gets denser in heavy quarks at chemical freeze-out as energy increases 
and so does the chance of formation of multiply heavy flavoured hadrons. Therefore, 
an enhanced production rate of these objects relative to singly heavy flavoured 
hadrons (like B's or D's) is distinctive of HIC and can be used as a probe of
thermalization of heavy quarks within the QGP, thence as a signal of QGP itself. 
This idea was advocated in refs.~\cite{thews,pbm} where the authors 
envisaged an enhancement of B$_c$ and $J/\psi$ mesons production from the QGP with 
respect to hadronic collisions at RHIC. In this paper, we amend and reinforce 
this picture by observing that the main advantage of baryons (and especially 
$\Xi_{bc}$ and $\Omega_{ccc}$ whose signal should be detectable at LHC) over 
quarkonia and B$_c$ is to enhance the difference from the ``background" of coherent 
production, i.e. direct production of these states from early hard scatterings, 
possibly followed by melting in the plasma. Furthermore, in view of this relatively 
large production, HIC become a suitable place to discover many of 
these yet unobserved states. 

That the production of multiply charmed baryons could be enhanced in high energy HIC 
was first proposed in ref.~\cite{levai} on the basis of a quark recombination model. 
A different coalescence picture was used in ref.~\cite{bielich}. However, in both
studies, no relation was set between production enhancement and heavy quark kinetic
equilibration in the plasma and quantitative predictions of production rates 
were dependent on an undetermined free parameter. 
On the contrary, predictions are definite within the statistical coalescence model 
(SCM), whose basic idea has been introduced in ref.~\cite{pbm} following a 
work on statistical production of $J/\psi$ \cite{gago}, and used thereafter by many 
authors \cite{vari}. In practice, the SCM is the statistical hadronization model
supplemented with the constraint of a fixed number of heavy quarks and antiquarks.

We are now going to calculate of the mean multiplicities $\langle n_j \rangle$ 
of heavy flavoured particles in the framework of the SCM. 
In HIC at high energy, electric charge, strangeness 
and baryon number conservation can be treated grand-canonically whereas charm and beauty
conservation cannot because the multiplicities of heavy flavoured hadrons are
not large. Thus, besides the number $N_c$ of c + $\cbar$ quarks and $N_b$ of 
b + $\bbar$ quarks, also the net charm $C$ and net beauty $B$ should be fixed. In 
fact, instead of $C$, $B$, $N_c$, $N_b$, any combination of these integers can 
be used to constrain the partition function. Indeed, it is advantegeous to use 
the numbers $\nuc$ of c, $\nucb$ of $\cbar$, $\nub$ of b and $\nubb$ of $\bbar$ 
quarks. The relevant partition function reads:
\begin{eqnarray}\label{partz}
  && Z(\nuc,\nucb,\nub,\nubb) = Z_l \left[ \prod_{f=c, \bar c, b, \bar b}
  \int_{-\pi}^\pi \frac{\d\phi_f}{2\pi} \e^{\i \nuf \phi_f} \right] \nonumber \\
 && \times \exp \Bigg[ \sum_j z_j \lambda_j \e^{-\i \nucj \phi_c - \i \nucbj 
   \bar\phi_c -\i \nubj \phi_b - \i \nubbj \bar\phi_b} \Bigg] 
\end{eqnarray}   
where $Z_l$ is the grand-canonical partition function including all light-flavoured
species; $z_j$ are the one-particle partition functions:
\begin{equation}\label{zeta}
  z_j = \frac{g_j V}{2\pi^2} m^2 T {\rm K}_2 \left(\frac{m}{T}\right) 
  \underset{m \gg T}{\simeq} g_j V \left(\frac{m T}{2 \pi}\right)^{3/2} 
  \e^{-m/T} ;
\end{equation}
$\lambda_j$ are the fugacities (with regard to electric, baryonic and strangeness 
charge) and $\nucj$, $\nucbj$, $\nubj$, $\nubbj$ are the number of c, $\cbar$, b, 
$\bbar$ quarks respectively of the of the $j^{\rm th}$ hadronic species; the 
factor $g_j$ in Eq. (\ref{zeta}) is its spin degeneracy and $T$ is the temperature. 
The sum over $j$ in 
Eq. (\ref{partz}) involves all hadrons containing heavy quarks. The primary average 
multiplicity of any of these, in events with fixed numbers of heavy quarks, reads:
\begin{equation}
  \langle n_j \rangle = z_j \lambda_j 
  \frac{Z(\nuc-\nucj,\nucb-\nucbj,\nub-\nubj,\nubb-\nubbj)}{Z(\nuc,\nucb,\nub,\nubb)}
\end{equation}
It can be realized that we would deal with factorized charm and bottom integrals 
in Eq. (\ref{partz}) were not for the presence of hadrons carrying both flavours, 
such as B$_c$. To recover factorization, we can expand the integrand in
Eq. (\ref{partz}) in power series of the $z_j$ of hadrons containing both c and b 
(anti-)quarks up to the first order. Defining:
\begin{eqnarray}\label{zetaf}
 {\hat Z_f}(\nuf,\nufb) = &&
 \int_{-\pi}^\pi \frac{\d\phi_f}{2\pi} \frac{\d\bar\phi_{\bar f}}{2\pi} \;
 \e^{\i \nuf \phi_f + \i \nufb \bar\phi_{\bar f}} \nonumber \\
 && \times \exp \Bigg[ \sum_{j_f}  z_{j_f} 
 \lambda_{j_f} \e^{-\i \nufj \phi_f -\i \nufbj \bar\phi_{\bar f}} \Bigg]
\end{eqnarray} 
where $f=c$ or $b$ and $j_f$ running over all hadrons containing $f$ or $\bar f$ quark, 
the partition function can be rewritten as:
\begin{eqnarray}\label{partz2}
\!\!\!\!\!\!\!\!&& Z(\nuc,\nucb,\nub,\nubb) \simeq Z_l 
 \Big[ \zhatc(\nuc,\nucb)\zhatb(\nub,\nubb)  + \sum_{j_{cb}} z_{j_{cb}} 
 \lambda_{j_{cb}} \nonumber \\
\!\!\!\!\!\!\!\!&& \times 
  \zhatc(\nuc-\nucj,\nucb-\nucbj) \zhatb(\nub-\nubj,\nubb-\nubbj) \Big]
\end{eqnarray}   
where $j_{cb}$ runs over all hadrons containing both flavours. It is
now clear from (\ref{partz2}) that we will henceforth be dealing with factorized 
canonical partition functions because both $\zhatc$ and $\zhatb$ do no longer 
involve hadrons with both flavours. Whether stopping the expansion in (\ref{partz2}) 
at first order is sufficient, will be discussed later. 
If we now denote by $a_{fn}$ the sum of $z_j \lambda_j$ for hadrons with $n$ 
units of open flavour $f$, $a_{\bar f n}$ for those with $n$ units of open anti-
flavour $\bar f$, and $a_{f0}$ for $f\bar f$ states, we can calculate the $\zhatf$ 
in (\ref{zetaf}) by expanding in powers of $a_{f0}$:   
\begin{eqnarray}\label{zexpand}
  && \zhatf(\nuf,\nufb) = 
  \int_{-\pi}^\pi \frac{\d\phi_f}{2\pi} \frac{\d\bar\phi_{\bar f}}{2\pi} \;
  \e^{\i \nuf \phi_f + \i \nufb \bar\phi_{\bar f}} \nonumber \\
  && \times \exp \Bigg[ 
  a_{f0} \e^{ - \i \phi_f -\i \bar\phi_{\bar f}} + \sum_{n=1}^3 a_{fn} 
  \e^{-\i n \phi_f} +  a_{\bar f n} \e^{ - \i n \bar\phi_{\bar f}} \Bigg] 
  \nonumber \\
  && = \sum_{h=0}^{\min(\nuf,\nufb)} 
  \frac{a_{f0}^h}{h!} \zeta(\nuf-h) \bar\zeta(\nufb-h)   
\end{eqnarray}
where:
\begin{equation}\label{greekz}
  \zeta(\nuf) \equiv  \int_{-\pi}^\pi \frac{\d\phi_f}{2\pi} \; \e^{\i \nuf \phi_f}
  \exp \Bigg[ \sum_{n=1}^3 a_{fn} \e^{-\i n \phi_f} \Bigg] 
\end{equation}
and similarly for $\bar\zeta(\nufb)$ with $a_{\bar f n}$ replacing the $a_{fn}$'s.
We note in passing that $\zeta(\nuf)=0$ if $\nuf<0$ and $\zeta(0)=1$.
The mass hyerarchy in heavy flavoured hadrons make $a_{f1} \gg a_{f2} \gg a_{f3}$ 
($f = c, b$), because, according to (\ref{zeta}), each 
term is suppressed with respect to the preceding one by a factor $\sim \exp(-m_f/T)$ 
where $m_f$ is the c or b quark mass. Neglecting $a_{f3}$ (i.e. $\Omega_{ccc}$ or
$\Omega_{bbb}$ baryons), the integrals in (\ref{greekz}) give rise to a polynomial 
expression:
\begin{equation}\label{poly}
 \zeta(\nuf) \simeq \sum_{k=0}^{[\nuf/2]} \frac{a_{f1}^{\nuf-2k}a_{f2}^k}
 {(\nuf-2k)!k!}
\end{equation}
and likewise for $\bar\zeta$. The ratio between the $k^{\rm th}$ and $(k-1)^{\rm th}$ 
term in the above polynomial is less than $(a_{f2}/a_{f1}^2) \nuf(\nuf-1)/k \equiv R_f/k$. 
Recalling the definition of the $a_{fn}$'s, one can roughly estimate it by assuming 
the mass of the hadrons to be the sum of the constituent masses of valence quarks and 
using Eq. (\ref{zeta}):
\begin{equation}\label{ratiof}
  R_f \equiv \frac{a_{f2}}{a_{f1}^2} \nuf(\nuf-1) 
 \approx \frac{\nuf(\nuf-1) \, \e^{2m_{u,d}/T}}
 {g_{\rm eff} V [\frac{(m_f + m_u)^2 T}{2\pi m_f}]^{3/2}} 
\end{equation}
where $g_{\rm eff}(T,\lambda)$ is an effective degeneracy parameter including the spin 
degeneracy and the different states with the same numbers of heavy quarks, weighted by 
the ratio $z_j \lambda_j /z_{1}\lambda_1$, $z_{1}$ being the one-particle partition function
of the lowest lying state. In Eq. (\ref{ratiof}) we tacitly 
assumed that $g_{\rm eff}$ is the same for hadrons with one or two heavy quarks, which 
approximately holds according to our numerical check. From known states with one heavy 
quark, either c or b, one expects $g_{\rm eff} \sim 10$ at $T=165$ MeV and $\lambda_j=1$. 
Taking constituent quark masses $m_c = 1.54$ GeV, $m_b = 4.95$ GeV, $m_{u,d} = 0.33$ GeV 
$m_s = 0.51$ GeV and $T = 165$ MeV, i.e. the fitted chemical freeze-out temperature 
at very large energy \cite{bronio,becahi3}, it turns out that $R_c \sim \nuc(\nuc-1)/0.34 
\; {\rm fm}^{-3} V$ and $R_b \sim \nub(\nub-1)/1.4 \; {\rm fm}^{-3} V $. Therefore, with 
the large volumes involved in HIC, $R_c, R_b$ are likely to be $\ll 1$ unless $\nuc$ or 
$\nub$ are consistently large. In this case we can approximate the $\zeta$ in 
Eq. (\ref{poly}) with its first term, i.e. 
$\zeta(\nuf) \simeq a_{f1}^{\nuf}/\nuf!$, so Eq. (\ref{zexpand}) becomes:
\begin{equation}\label{zexpand2}
  \zhatf(\nuf,\nufb) \simeq \sum_{h=0}^{\nuf} \frac{a_{f0}^h}{h!} 
  \frac{a_{f1}^{\nuf-h}}{(\nuf-h)!}\frac{a_{\bar f 1}^{\nufb-h}}{(\nufb-h)!}
\end{equation} 
Again, the ratio between the $h^{\rm th}$ and the $(h-1)^{\rm th}$ term is less 
than $a_{f0}\nuf\nufb /a_{f1} a_{\bar f 1} h$, which is approximately equal to $R_f/h$,
with $R_f$ quoted in Eq. (\ref{ratiof}), because $\nuf$ differs from $\nufb$ by few 
units and being $a_{f1}$ very close to $a_{\bar f 1}$ if chemical potentials 
are not too large. Therefore, under the same conditions needed for its validity, the 
sum in (\ref{zexpand2}) can be approximated with its
first term, i.e. $\zhatf(\nuf,\nufb) \simeq a_{f1}^{\nuf} a_{\bar f 1}^{\nufb}
/ \nuf! \nufb!$. By using this approximation, we can finally estimate the ratio 
between the first and the zeroth order terms in the expansion (\ref{partz2}). 
For instance, for B$_c$ mesons carrying one c and one $\bbar$ quark:
\begin{eqnarray}\label{rcb}
 && \sum_{j_{cb}} z_{j_{cb}} \lambda_{j_{cb}} \frac{\zhatc(\nuc-1,\nucb) 
  \zhatb(\nub,\nubb-1)}{\zhatc(\nuc,\nucb)\zhatb(\nub,\nubb)} \nonumber \\
  &\simeq& \sum_{j_{cb}} z_{j_{cb}} \frac{\nuc \nubb}{a_{c1} a_{\bar b 1}} \approx 
  \frac{\nuc \nubb \, \e^{2m_{u,d}/T}}{g_{\rm eff} V [\frac{m_c m_b T}
  {2 \pi (m_c+m_b)}]^{3/2}} \equiv R_{cb}
\end{eqnarray}
For hadrons with two b's and one c and viceversa, it can be easily shown that this 
ratio is even smaller. If $R_{cb} \ll 1$ (which is likely to be), the first order 
term in the expansion of the partition function (\ref{partz2}) is negligible. Under 
these circumstances, and provided that the aforementioned conditions on $R_c$, $R_b$
are met, the primary average multiplicity of heavy flavoured hadrons for fixed 
number of c, $\cbar$, b, $\bbar$ quarks is especially simple:
\begin{equation}\label{fixed} 
 \langle n_j \rangle = z_j \lambda_j \!\!\!\prod_{f=c, \bar c, b, \bar b, \nu_{fj}>0}
     \!\!\!\!\!\!\! \frac{\nu_f(\nu_f-1)\ldots(\nu_f-\nu_{fj}+1)}{a_{f1}^{\nu_{fj}}}
\end{equation}
The (\ref{fixed}) is to be further averaged over the multiplicity distribution $p_{\nuc}$
of $\ccb$ and $p_{\nub}$ of $\bbb$ pairs created in a single collision. If they are 
independently produced, this is a Poisson distribution and, for open flavoured hadrons, 
the sum (\ref{fixed}) yields its factorial moments, i.e.:  
\begin{equation}\label{final}
     \langle\langle n_j \rangle\rangle = z_j \lambda_j\!\!\! \prod_{f=c, \bar c, b, \bar b}
  \!\!\!   \left(\frac{\langle \nu_f \rangle}{a_{f1}}\right)^{\nu_{fj}} \equiv 
     z_j \lambda_j\!\!\! \prod_{f=c, \bar c, b, \bar b} \eta_f^{\nu_{fj}} 
\end{equation}
whereas for quarkonia it is more complicated since $\nuc=\nucb$ and $\nub=\nubb$.
Eq. (\ref{final}) is our final formula. As has been mentioned, it is an approximated
expression valid if $R_c, R_b, R_{cb} \ll 1$. However, it can be shown, by performing
an asymptotic expansion of the integral (\ref{zexpand}) \cite{prepa} that it still
holds under the weaker condition $R_f/\nuf \ll 1$ if $\nuf$ 
is large. Altogether, the Eq. (\ref{final}) implies that the contribution of hadrons 
carrying more than one heavy flavoured quark in the balance equations 
$\sum_j \langle n_j \rangle \nufj=\langle \nuf \rangle$ is neglected. It is interesting 
to note that the enhancement factors 
$\eta_f=\langle \nu_f \rangle/a_{f1}$ are proportional to the {\em density} of heavy 
quarks at the hadronization temperature, as $a_{f1} \propto V$. 
Therefore, the ratio between multiply and singly heavy flavoured hadrons,
proportional to $(\langle \nuf \rangle/V)^{\nu_{fj}-1}$, increases with centre-of-mass 
energy because the volume (or the charged multiplicity) increases much slower than 
$\scc$ and $\sbb$ do.

The formula (\ref{final}) can now be applied to estimate the average multiplicity of 
multiply heavy flavoured hadrons in HIC at RHIC and LHC. For sake of simplicity, we 
will confine ourselves to full phase space integrated quantities, disregarding spectra. 
To get started, we need the cross sections $\scc$ and $\sbb$ in pp collisions at 
relevant energies. There is a large uncertainty on these values; recent calculations 
indicate $\scc = 110 - 656 \; \mu{\rm b}$ and $\sbb = 1.2 - 2.86 \; 
\mu{\rm b}$ at $\sqrt s = 200$ GeV \cite{cacciari} and $\scc = 3.4 - 9.2 \; 
{\rm mb}$ and $\sbb = 88 - 260 \; \mu{\rm b}$ at $\sqrt s = 5.5$ TeV \cite{review}. 
The production of heavy quark pairs is a hard process and should scale like the 
number of collisions $N_{\rm coll}$ in the Glauber model. Specifically, if 
$\sigma_{\rm inel}$ is the total inelastic NN cross section, the average multiplicity 
of $\ccb$ pairs is 
$\langle \nuf \rangle = \langle \nufb \rangle = N_{\rm coll} \sff/\sigma_{\rm inel}$. 
At RHIC, in Au-Au collisions at $\sqrt s_{NN} = 200$ GeV, $\sigma_{\rm inel} \simeq 
42$ mb and for a 5.5\% centrality selected sample, corresponding to an impact parameter 
range 0-3.5 fm, $N_{\rm coll} = 1080$ \cite{web}. Thus, the average 
multiplicity of $\ccb$ pairs ranges from 2.8 to 17, whereas for $\bbb$ pairs from 
0.03 to 0.07. On the other hand, at LHC, in Pb-Pb collisions at $\sqrt s_{NN} = 5.5$ 
TeV, $\sigma_{\rm inel} \simeq 60$ mb and for a 5.1\% centrality selected sample, 
corresponding to the same impact parameter above, $N_{\rm coll} = 
1670$ \cite{web}. In this latter case, the average multiplicity of $\ccb$ pairs 
ranges from 95 to 256 and from 2.4 to 7.2 for $\bbb$ pairs. 
It should be noted that these estimates do not take into account possible structure
function saturation effects, which are predicted to reduce heavy quark cross section 
at LHC \cite{kharzeev}. Since there is not a clearcut evidence of this phenomenon as
yet, we stick to the traditional picture of $N_{\rm coll}$ scaling, though this 
possibility is worthy of consideration in the future.  

The other key ingredient in our calculation is the volume $V$. 
In order to extrapolate from SPS to RHIC, we take advantage of the fact that 
$V$ is proportional to the average multiplicity of pions. In Pb-Pb at SPS at 
$\sqrt s_{NN} = 17.2$ GeV, 
$V \simeq 3.5 \,10^3$ fm$^3$ in full phase space \cite{becahi3}, and $\langle \pi^+ + 
\pi^- \rangle \simeq 1258$ \cite{na49}. At RHIC, at $\sqrt s_{NN} = 200$ GeV in full 
phase space $\langle \pi^+ + \pi^- \rangle \simeq 3343$ \cite{brahms} leading to 
$V \approx 10^4$ fm$^3$. This extrapolation assumes very little variation of temperature 
and baryon-chemical potentials from SPS to RHIC, which approximately holds 
\cite{becahi3}. 
In order to extrapolate from RHIC to LHC, we pragmatically use the saturation model 
(which proved to be successful in extrapolating multiplicity from SPS to RHIC)
which predicts an increase in $\langle n_{\rm ch} \rangle$ by a factor $\simeq 4.5$ 
from $\sqrt s_{NN} = 200$ GeV to 5.5 TeV \cite{nardi}. This means that 
$V_{LHC} \approx 4.5\,10^4$ fm$^3$ with a fair uncertainty up to a factor 2. 
By using the above values for $V$, and, conservatively, the upper estimates 
for $\langle \nuc \rangle$ and $\langle \nub \rangle$ we obtain, from Eqs. (\ref{ratiof})
and (\ref{rcb}) $R_c = {\mathcal O}(10^{-2}), R_b= {\mathcal O}(10^{-7}), R_{cb} = 
{\mathcal O}(10^{-4})$ at RHIC and $R_c = {\mathcal O}(1), R_b= {\mathcal O}(10^{-4}), 
R_{cb} = {\mathcal O}(10^{-1})$ at LHC, by using as input parameters $m_c,m_b,m_{u,d}$
and $T$ the same quoted below Eq. (\ref{ratiof}). Therefore, both at RHIC and LHC 
energies the formula (\ref{final}) should be fairly accurate, because either the $R$'s
are $\ll 1$ or, like in the charm sector at LHC, $\langle \nuc \rangle \gg 1$ and
$R_c/\langle \nuc \rangle \ll 1$.

We can now perform our predictions. To estimate the $\eta_f$'s, we use the approximation 
$a_{f1} \simeq g_{\rm eff} V [(m_f + m_u)T/2\pi]^{3/2} \exp[(-m_f+m_u)/T]$ with input 
parameters like for the ratios $R_f$. We then obtain $\eta_c \approx 1.7 - 10$ and 
$\eta_b \approx (3.5 - 8.2) \cdot 10^6$ at RHIC and $\eta_c \approx 12 - 34$ and 
$\eta_b \approx (0.63 - 1.9) \cdot 10^8$ at LHC.
With these numbers, assuming the mass of multiply heavy flavoured hadrons to be
the sum of its quarks constituent masses, using Eq. (\ref{final}) with $\lambda_j=1$
(i.e. taking vanishing chemical potentials) and appropriate spins, we get average 
{\em primary} yields of doubly charmed baryons (like $\Xi_{cc}$ and $\Omega_{cc}$) 
between 
$0.7 \cdot 10^{-4}$ and $7 \cdot 10^{-3}$ in central collisions at RHIC and between 
0.019 and 0.38 at LHC. For mixed charmed-beautiful hadrons (like $\Xi_{bc}$, 
$\Omega_{bc}$ and $B_c$ meson), the yields should range between $4 \cdot 10^{-7}$ and 
$6 \cdot 10^{-5}$ at RHIC and $3 \cdot 10^{-4}$ and $0.022$ at LHC. For 
doubly beautiful baryons (like $\Xi_{bb}$ and $\Omega_{bb}$), our estimates range 
between $2 \cdot 10^{-9}$ and $3 \cdot 10^{-8}$ at RHIC and between 
$2.6 \cdot 10^{-6}$ and $7 \cdot 10^{-5}$ at LHC.
For the $\Omega_{ccc}$ baryon, the predicted yields are affected by a large uncertainty 
due to the cubic dependence on $\eta_c$; they range between $7 \cdot 10^{-7}$ and $10^{-4}$
at RHIC and between $10^{-3}$ and $0.03$ at LHC. For charmed baryon yields 
at LHC, the predictions of the Eq. (\ref{final}) turn out to be in good agreement with 
a preliminary calculation with the exact formula. All of the previous yields are 
enhanced by the feeding from heavier states, even by factor of about 4-5; another 
factor $\approx 2$ comes from antiparticle yields. These factors should roughly 
compensate for the limited rapidity window accessible to experiments. 
Therefore, while at RHIC only doubly charmed hadrons seem to be
within reach, at LHC, with a statistics of $10^7$ central events/year, doubly and
triply charmed, charmed-beautiful, and perhaps doubly beautiful hadrons 
could in principle be observed.

We can now compare the above yields with the predictions by production models 
based on QCD hard scattering. At LHC, a model where heavy diquarks produced in 
${\cal O}(\alpha_S^4)$ diagrams are assumed to fully hadronize into ccq-baryons, yields 
an upper bound on inclusive production of $(10^{-4}-10^{-3}) \langle \nuc \rangle$
at $\sqrt s = 14$ TeV \cite{kise} in pp, to be compared with $g_{\rm eff} (0.8 -2) \cdot
10^{-3} \langle \nuc \rangle$ from coalescence in HIC at $\sqrt s_{NN} = 5.5$ TeV. 
A larger difference is found in the $\Xi_{bc}$ sector, where in $\ppb$ at 
$\sqrt s = 1.8$ TeV the 1S-wave cross-section is predicted to be around 1 nb 
\cite{kise2}, implying a ratio $\langle \Xi_{bc} \rangle/\langle \nub \rangle \sim 10^{-5}$
to be compared with $g_{\rm eff}^{1S} (3-9) \cdot 10^{-4}$ 
from coalescence, i.e. at least one order of magnitude larger. Since the production
process is ${\cal O}(\alpha_S^6)$, the difference is even larger for $\Omega_{ccc}$, 
for which a recent calculation \cite{iran} predicts a ratio 
$\langle \Omega_{ccc}\rangle /\langle \nuc \rangle = {\cal O}(10^{-7})$
in pp at $\sqrt s = 14$ TeV; this is between 2-3 orders of magnitude lower than our 
estimated ratio from coalescence at $\sqrt s_{NN} = 5.5$ TeV i.e. $(0.1-1)\cdot 10^{-4}$.
For charmonia and B$_c$, the two mechanisms give closer predictions. For $\langle
J/\psi \rangle/\langle \nuc \rangle$, the difference is estimated to be a factor 
about 2.5 at LHC \cite{review}, whilst for B$_c$ cross sections calculations 
\cite{kisebc} at $\sqrt s = 14$ TeV imply a ratio $\langle {\rm B}_c 
\rangle / \langle \nub \rangle = {\cal O}(10^{-3})$, around the same as from 
coalescence at $\sqrt s_{NN} = 5.5$ TeV. Also, it should be pointed out that,
unlike $J/\psi$ and B$_c$, multiply heavy flavoured hadrons have not 
been measured in hadronic collisions, so the predictions of the models based on hard 
scattering are still to be checked. 

The predominant uncertainty on the previous estimates is that on heavy quark 
cross section. Other relevant uncertainties are the those on masses, effective 
degeneracy, extrapolated temperature and charged multiplicities, a modulation of 
the production as a function of rapidity as well as the replacement of the approximated 
formula (\ref{final}) with the exact one. Yet, all these effects, which will be 
discussed in detail in a forthcoming paper \cite{prepa}, cannot alter 
the ratios of multiply to singly flavoured hadron yields by one order of magnitude. 
So the conclusion remains that 
if statistical coalescence scheme applies, a large enhancement in the measurement
of $\langle \Xi_{bc} \rangle / \langle B \rangle$, which becomes dramatic for 
$\langle \Omega_{ccc} \rangle / \langle D \rangle$, could be found in heavy ion 
collisions with respect to pp at the LHC energy. This could be a clear indication 
of QGP formation.




\end{document}